\documentclass[twocolumn]{aims}
\usepackage{multirow}
\usepackage{array}
\usepackage{txfonts}
\usepackage{cite}
\def\typeofarticle{Research article}
\def\currentvolume{3}
\def\currentissue{x}
\def\currentyear{2024}

\def\ppages{xxx--xxx}
\def\DOI{10.3934/mmc.2025xxx}
\def\Received{July 2025}
\def\Revised{August 2025}
\def\Accepted{September 2025}
\def\Published{December 2025}

\numberwithin{equation}{section}

\makeatletter
\renewcommand{\@biblabel}[1]{#1\hfill \hspace{-0.2cm}}
\makeatother

\newtheorem{remark}{Remark}[section]

\begin{document}

\title{Applications of structural equation modeling and mathematical statistics to the triggering mechanism of a class of liquors consumer behaviors in Sichuan province}

\author{%
  Ruofeng Rao\affil{1,2}\corrauth}

% \shortauthors is used in copyright information in the end of the paper
\shortauthors{the Author(s)}

\address{%
  \addr{\affilnum{1}}{School  of Mathematics, Chengdu Normal University, Chengdu  611130, China}
  \addr{\affilnum{2}}{Faculty of Management, Shinawatra University, Pathum Thani 12160, Thailand}  }

% corresponding author
\corraddr{Email: ruofengrao@163.com; Tel: +86-028-66775259; Fax: +86-028-66772000.
}

\begin{abstract}
Structural Equation Modeling (SEM) systematically validated hierarchical pathways among multiple factors by constructing a dual framework integrating latent variable measurement and path analysis, utilizing covariance matrices derived from online questionnaires of Wuliangye consumers in Sichuan Province. Statistical analysis quantified path coefficient significance through maximum likelihood estimation, revealing via factor loadings and goodness-of-fit tests that consumer ethnocentrism directly promotes purchase intention, while simultaneously refuting the null hypothesis regarding perceived behavioral control-thus deconstructing the "trigger-transmission" causal chain among variables. Crucially, SEM findings revealed environmental stimuli as the predominant factor, indirectly influencing purchasing behavior through perceived value, contrary to existing literature asserting equal impacts from consumer ethnocentrism, environmental stimuli, and perceived behavioral control. Statistical evidence further demonstrated higher online purchase frequency for premium Wuliangye liquor, aligning with Generation Z's e-commerce preferences. By implementing stricter website-based participant screening than prior studies, this research optimized the analytical model, yielding data-driven strategic recommendations: strengthening e-commerce platforms, enhancing promotional expertise, leveraging cultural localization, and prioritizing premium product development. These actionable insights significantly advance sales optimization strategies for Wuliangye products in Sichuan's dynamic market.

\end{abstract}


\maketitle

\section{Introduction}
This section begins by explaining why this study employs Structural Equation Modeling (SEM) combined with questionnaire survey data to investigate the drivers of consumer purchase behavior and how these factors trigger purchase actions, rather than relying solely on SPSS for analysis with the same survey data.
In fact,
SEM is employed in this study to investigate the complex drivers of consumer purchase behavior, surpassing the capabilities of traditional SPSS-based analyses (e.g., correlation, regression) for several key reasons. Consumer behavior involves abstract constructs (latent variables) like perceived value, attitude, and intention, which SEM rigorously models by simultaneously evaluating both the measurement model (assessing how survey items reflect these latent constructs and accounting for measurement error) and the structural model (testing hypothesized causal pathways). Crucially, SEM excels at analyzing intricate networks of direct and indirect effects (e.g., mediation), allowing researchers to delineate precisely how multiple antecedents (e.g., price sensitivity, social influence) interact and propagate through pathways (e.g., attitude mediating the effect of brand awareness on purchase) to ultimately trigger the purchase decision. Furthermore, SEM provides holistic assessment through model fit indices, evaluating the plausibility of the entire proposed theoretical framework depicting the drivers and their interrelationships. This integrated approach offers a more comprehensive, accurate, and theoretically grounded framework for mapping the multifaceted causal mechanisms underlying consumer purchases than methods focusing solely on observed variables and isolated relationships.
Key methodological advantages include robust error variance control, differentiation of causal pathways (direct vs. mediated effects), and capacity for cross-population analyses (e.g., demographic subgroups), making it particularly valuable for generating evidence-based strategic insights (\cite{authour1,authour2,authour3}). SEM's analytical rigor stems from its ability to isolate mediation mechanisms (e.g., mapping price elasticity impacts across supply chain tiers) while accounting for measurement variability, thereby yielding actionable intelligence for product portfolio optimization and channel management (\cite{authour4,authour5,authour6}).

Consequently, numerous relevant studies have investigated the triggering mechanisms of consumer purchase behavior utilizing questionnaire surveys and Structural Equation Modeling (SEM). For instance, Maksan, Damir, and Marija \cite{authour7} examined the triggering mechanism of Croatian wine purchase behavior based on the Extended Theory of Planned Behavior (ETPB). Similarly, Molinillo, Aguilar-Illescas, Anaya-Sš¢nchez, and LišŠbana-Cabanillas \cite{authour8} applied the Stimulus-Organism-Response (SOR) theory to investigate the triggering mechanism of consumers' online purchase behavior. More recently, Rao and Photchanachan \cite{authour9} explored the triggering factors and mechanisms of Wuliangye consumer purchase behavior in Sichuan Province, China, using a hybrid theoretical framework integrating both the Extended Theory of Planned Behavior (ETPB) and the Stimulus-Organism-Response (SOR) theory. Collectively, these investigations into purchase intention and purchase behavior theories are fundamentally grounded in questionnaire survey methodology and Structural Equation Modeling.

Next, I discuss why, despite existing literature (\cite{authour9}) examining Wuliangye's marketing strategies, this paper still investigates the factors influencing Wuliangye consumers' purchase behavior. This is because the survey participants in Reference \cite{authour9} were individuals who had lived in Sichuan Province for over three years, but they might not currently reside there. Therefore, this online survey specifically targets individuals who are currently living in Sichuan Province. Furthermore, the analysis of this survey data  revealed a new research model, which differs from the one in Reference \cite{authour9}. Crucially, this new model provides actionable insights for genuinely improving Wuliangye's sales within Sichuan Province.

This paper makes the following novel contributions:

a) Utilizing rigorously screened survey data, I developed a new research model. By running Amos structural equation modeling (SEM), I identified the trigger mechanisms for Wuliangye consumers' purchasing behavior in Sichuan Province. This provides actionable recommendations for tangibly improving sales of Wuliangye's product line within the province.

b) My research model integrates the Extended Theory of Planned Behavior (ETPB) from Reference \cite{authour7} and the Stimulus-Organism-Response (SOR) theory from Reference \cite{authour8} into a cohesive new structural equation model. This integrated model demonstrated good statistical fit and contributes significant new knowledge.   From a methodological standpoint, while Structural Equation Modeling (SEM) is a well-established technique, the innovation of this work does not stem from the technique itself but from its application to a novel theoretical framework and a unique empirical context. The methodological contribution is manifested in: (1) The development and empirical validation of a new integrated ETPB-SOR theoretical model, as previously described; (2) The specific operationalization of constructs (e.g., defining 'Environmental Stimulus' in the context of Wuliangye's marketing efforts and online platform) tailored to the Chinese premium liquor market; and (3) The rigorous testing of this model on a unique and strategically important sample-current residents of Sichuan Province, the core market for Wuliangye. This approach follows the precedent set by similar studies that apply robust methodological tools like SEM to test new conceptual models in specific contexts, thereby generating novel insights \cite{authour10,authour11}. The value lies not in inventing a new statistical method, but in leveraging a powerful method to answer new research questions and test new conceptual relationships within a clearly defined and understudied setting.  This integrated model not only demonstrated good statistical fit but also contributes significant new knowledge to the theoretical landscape of consumer behavior research .

 c) In comparison to recent studies applying SEM in consumer behavior research (\cite{authour12,authour13,authour14}), my integrated ETPB-SOR model offers a more nuanced understanding of regional premium liquor purchases by simultaneously accounting for internal cognitive mechanisms (via ETPB) and external market stimuli (via SOR). Unlike \cite{authour12}, which focused on general e-commerce contexts, or \cite{authour13} and \cite{authour14}, which examined broad consumer segments without regional specificity, my model is tailored to the socio-cultural and economic context of Sichuan Province, providing localized and actionable strategic insights. Furthermore, whereas \cite{authour13} and \cite{authour14} relied on simpler theoretical frameworks, my hybrid approach captures both direct and mediated effects, offering a more comprehensive causal mapping of purchase behavior drivers.

  Although the empirical context is region-specific, the integrated ETPB-SOR theoretical framework is generalizable to other regional markets or product categories, as it captures universal cognitive and environmental mechanisms that drive consumer behavior \cite{authour15}. Thus, while the strategic insights are tailored to Sichuan and Wuliangye, the methodological and theoretical contributions offer broader applicability.

  d) Theoretical Contribution: This study makes a significant contribution to academia by proposing and empirically validating a novel hybrid theoretical framework that integrates the Extended Theory of Planned Behavior (ETPB) and the Stimulus-Organism-Response (SOR) model. While both theories have been used independently, their synthesis in the context of regional premium liquor consumption is novel. This research provides a more holistic and nuanced lens to decipher the complex interplay between internal cognitive mechanisms (e.g., ethnocentrism, perceived control) and external market stimuli in driving purchase behavior, thereby addressing a gap in the existing literature.

e) Methodological Contribution: The study offers a robust methodological demonstration of employing Structural Equation Modeling (SEM) to test complex, theory-driven models with mediating effects in a specific regional and product context. It provides a replicable blueprint for future research aiming to understand culturally-specific consumer behavior.

f) Empirical Contribution: The findings offer unique empirical insights into the consumer behavior dynamics within the Chinese premium liquor market, specifically in Sichuan Province-a crucial yet under explored context. The identification of environmental stimuli as the dominant factor mediated by perceived value, alongside the nuanced roles of ethnocentrism and perceived behavioral control, delivers concrete, actionable knowledge that enriches the academic discourse on regional marketing and consumer decision-making.

\section{Preliminaries-Mathematical modeling based on consumer behavior theories and statistical data analysis}
\subsection {From linear regression to structural equation modeling}
To establish a rigorous mathematical foundation for the Structural Equation Model (SEM) used in this study, I begin with the classical linear regression model, which serves as a building block for more complex latent variable models. Consider the multiple linear regression model:
$$y=X\beta+\varepsilon,$$
 where $y$ is an $n\times 1$ vector of observed responses, $X$ is an $n\times p$ matrix of predictors, $\beta$ is a
$p\times 1$ vector of regression coefficients, and $\varepsilon$ is an $n\times 1$ vector of errors, typically assumed to be normally distributed with mean zero and constant variance $ \sigma^2$. The ordinary least squares (OLS) estimator minimizes the residual sum of squares:
$$\hat{\beta}=\text{arg} \min\limits_\beta\|y-X\beta\|^2=(X^TX)^{-1}X^Ty.$$
 While OLS is optimal under Gauss-Markov assumptions, it cannot handle latent constructs or measurement error in predictors--a key limitation in behavioral research. This motivates the use of factor analysis, which models observed variables
$x$ as linear functions of latent factors
$\xi$:
$$x=\Lambda_x\xi+\delta,$$
where $\Lambda_x$  is a matrix of factor loadings and $\delta$ is a vector of measurement errors. Combining this measurement model with a structural model that specifies relationships among latent variables leads to the full SEM framework.

The general SEM consists of two parts: the measurement model and the structural model. The measurement model for exogenous and endogenous variables is given by:

$$ x=\Lambda_x\xi+\delta,\qquad\,y=\Lambda_y\eta+\epsilon,$$
 where $x$ and $y$ are vectors of observed exogenous and endogenous variables, respectively; $\Lambda_x$
  and $\Lambda_y$ are matrices of factor loadings; $\xi$ and $\eta$ are latent exogenous and endogenous variables; and
$\delta$ and $\epsilon$ are measurement errors.  The structural model specifies the causal relationships among latent variables:

$$\eta=B\eta+\Gamma\xi+\zeta,  $$
 where $B$ and $\Gamma$ are matrices of path coefficients, and
$\zeta$ represents the structural disturbance term. The model assumes $E(\zeta)=0,\,\text{Cov}(\xi,\zeta)=0,$ and that $I-B$ is
  invertible.

The covariance structure of the observed variables is derived from the model parameters. Let $\Phi=\text{Cov}(\xi)$
 and $\Psi=\text{Cov}(\zeta).$ Then the implied covariance matrix
$\sum(\theta)$ is:
\begin{small}
$$\sum(\theta)=\begin{pmatrix}  \Omega^*&
\Lambda_y(I-B)^{-1}\Gamma\Phi\Lambda_x^T      \\
\Lambda_x\Phi\Gamma^T[(I-B)^{-1}]^T\Lambda_y^T  &\Lambda_x\Phi\Lambda_y+\Theta_\delta   \end{pmatrix},$$
\end{small}
where $\Omega^*= \Lambda_y(I-B)^{-1}(\Gamma\Phi\Gamma^T+\Psi) [(I-B)^{-1}]^T\Lambda_y^T+\Theta_\epsilon $ while $\Theta_\epsilon$ and $\Theta_\delta$ are covariance matrices of measurement errors. Model parameters are estimated by minimizing the discrepancy between the sample covariance matrix $S$ and $\sum(\theta)$, e.g., via Maximum Likelihood (ML):
$$ F_{ML}=\log|\sum(\theta)|+\text{tr}(S{\sum}^{-1}(\theta))-\log|S|-p,$$
which is derived from the multivariate normal log-likelihood function.

 This SEM framework generalizes and unifies several multivariate techniques¡ªincluding regression, factor analysis, and path analysis¡ªinto a single, flexible modeling system capable of testing complex theories with latent variables and mediated effects.

Inspired by References \cite{authour7,authour8,authour9}, this study integrates the Extended Theory of Planned Behavior (ETPB) and the Stimulus-Organism-Response (SOR) theory. Combined with the analysis of my survey data, I developed a new structural equation model for this research (see Figure 1).

Based on the research model depicted in Figure 1, the structural equations for the latent variables are formulated as follows. Let $\xi=[ConsEth, EnvSt,PBC]^T$
   denote the vector of exogenous latent variables, and
$\eta=[PerVa,PB]^T$
  the vector of endogenous latent variables. The structural model is then:
     $$\begin{pmatrix}   \eta_1\\ \eta_2  \end{pmatrix}=\begin{pmatrix}   0& 0\\ \beta_{21}&0 \end{pmatrix}
\begin{pmatrix}   \eta_1\\ \eta_2  \end{pmatrix}+\begin{pmatrix}  \gamma_{11}& \gamma_{12}&\gamma_{13} \\
\gamma_{21}&\gamma_{22}&\gamma_{23} \end{pmatrix}\begin{pmatrix} \xi_1\\ \xi_2\\ \xi_3\end{pmatrix}+\begin{pmatrix}  \zeta_1\\ \zeta_2\end{pmatrix},\eqno(2.1)$$
    where $\eta_1$
  represents Perceived Value (PerVa),
$\eta_2$
  represents Purchase Behavior (PB), $\xi_1$
  represents Consumer Ethnocentrism (ConsEth),
$\xi_2$  represents Environmental Stimulus (EnvSt), and
$\xi_3$ represents Perceive Behavioral Control (PBC). The coefficients $\gamma_{ij}$ and $\beta_{21}$
are the path coefficients to be estimated, and $\zeta_1$ and $\zeta_2$
  are disturbance terms. This formulation captures the direct and mediated effects hypothesized in H1-H6, aligning with the SEM framework introduced earlier.

\subsection{Methodological innovation of the proposed SEM}
The innovation of this study lies not only in the novel integration of ETPB and SOR theories but also in the formalization and identification of a hybrid latent variable model that captures both internal cognitive processes and external market stimuli. Unlike traditional regression models, the proposed SEM explicitly accounts for measurement error and allows for the simultaneous estimation of direct and indirect (mediated) effects. For instance, the mediation hypothesis H5: EnvSt$\to$ PerVa$\to$ PB
 is formally tested via the product of paths $\Gamma_{\text{EnvSt}\to\text{PerVa}}\cdot B_{\text{PerVa}\to \text{PB}}$,
 with standard errors derived using the delta method or bootstrap:

 $$\sigma_{\text{indirect}}^2\approx\gamma^2\sigma_B^2+B^2\sigma_\gamma^2+\sigma_\gamma^2\sigma_B^2,$$
  where $\gamma$ and $B$ are the path coefficients, and $\sigma_\gamma^2, \,\sigma_B^2$
 their variances. This enables a more precise decomposition of total effects into direct and indirect components, as shown in Tables 12-14.

Moreover, the model's identification is ensured by constraining sufficient parameters and using multiple indicators per latent construct. The overall fit is assessed using a suite of indices (e.g., RMSEA, CFI, TLI), which are functions of the minimized discrepancy
$F_{ML}$ and degrees of freedom. The rigorous application of this SEM methodology to a region-specific consumer context represents a significant advancement over prior studies that relied on simpler statistical techniques or less integrated theoretical frameworks.

\subsection{Methodological innovation and statistical data analysis endow the proposed research model with novelty}
The methodological innovations detailed in Section 2.2, combined with rigorous statistical data analysis, fundamentally establish the novelty and superiority of the proposed research model (Figure 1). Firstly, the innovative integration of ETPB and SOR theories into a cohesive hybrid model, as visualized in Figure 1, provides a more comprehensive and nuanced theoretical lens than the individual models presented in Figure 2 (ETPB-only from \cite{authour7}) and Figure 3 (SOR-only from \cite{authour8}). This integration allows for the simultaneous examination of internal cognitive mechanisms (e.g., Consumer Ethnocentrism, Perceived Behavioral Control) and external market stimuli (Environmental Stimulus), mediated by Perceived Value, enabling a deeper causal analysis of purchase behavior drivers specific to regional premium liquor consumption.

Secondly, the stringent data collection criteria¡ªspecifically, the online survey (ID: 284362158) conducted via the Wenjuanxing platform restricted to participants with IP addresses verified to be within Sichuan Province¡ªensures that the sample consists of current residents of the province. This is a critical improvement over Reference \cite{authour9}, which surveyed individuals who had lived in Sichuan for over three years but might not currently reside there. This fundamental difference in sampling frames means the data underlying Figure 1 are more relevant, timely, and geographically precise for analyzing in-situ purchase behavior, directly contributing to the model's uniqueness and actionable insights for Sichuan-specific marketing strategies.

Thirdly, the preliminary statistical analyses¡ªincluding reliability tests (Cronbach¡¯s ŠÁ), validity assessments (KMO, Bartlett¡¯s test), Exploratory Factor Analysis (EFA), and Confirmatory Factor Analysis (CFA)¡ªplayed a decisive role in refining the proposed model. These analyses, summarized in Tables 3-8, confirmed the robustness, convergent validity, and discriminant validity of the constructs. Consequently, this data-driven refinement process led to the exclusion of certain factors present in the model of Reference \cite{authour9} (Figure 4), such as 'Attitudes,' which were found to be less critical or redundant in the specific context of Wuliangye consumers in Sichuan. The elimination of these extraneous variables results in a more parsimonious, focused, and powerful model (Figure 1) that directly captures the core triggers of purchase behavior¡ªConsumer Ethnocentrism, Environmental Stimulus, Perceived Behavioral Control, and their mediation through Perceived Value.

In summary, the synergistic combination of methodological innovation (the hybrid ETPB-SOR SEM framework) and rigorous, context-specific statistical data analysis (using a current resident sample and psychometric validation) renders the proposed research model (Figure 1) truly unique and superior. It not only differs fundamentally from prior models (Figures 2, 3, and 4) but is also optimally tailored to decipher the purchase behavior triggers of Wuliangye consumers in Sichuan Province, thereby providing a solid foundation for formulating precise and effective sales strategies within the region.

 While prior research has applied ETPB \cite{authour7} and SOR \cite{authour8} in isolation, the key innovation of this study lies in the novel integration of these two theoretical frameworks to form a cohesive hybrid model specifically tailored to investigate the purchase behavior of Wuliangye consumers in Sichuan Province. This integration is not merely additive but synergistic: the ETPB components (e.g., Consumer Ethnocentrism, Perceived Behavioral Control) capture the internal, decision-making mechanisms of consumers, while the SOR components (Environmental Stimulus, Perceived Value) account for the external, market-driven stimuli and the cognitive-affective processing they trigger. This hybrid approach provides a more comprehensive and nuanced theoretical lens than either theory could offer alone, specifically designed to decipher the complex drivers of premium liquor purchases in a distinctive regional market. Furthermore, unlike the model in \cite{authour9} which targeted former residents of Sichuan, my model is rigorously derived from and validated against data from current residents, ensuring its relevance and actionable insights for marketing strategies within the province.

   Based on the structural equation modeling framework mentioned above, this study defines the following latent variables and constructs research hypotheses.
The model incorporates the following latent constructs: Consumer Ethnocentrism (ConsEth), Environmental Stimulus (EnvSt), Perceived Behavioral Control (PBC), Perceived Value (PerVa), and Purchase Behavior (PB). The hypothesized paths are illustrated in Figure 1.

\begin{figure}[!htbp]
\centering
\includegraphics[width=0.45\textwidth]{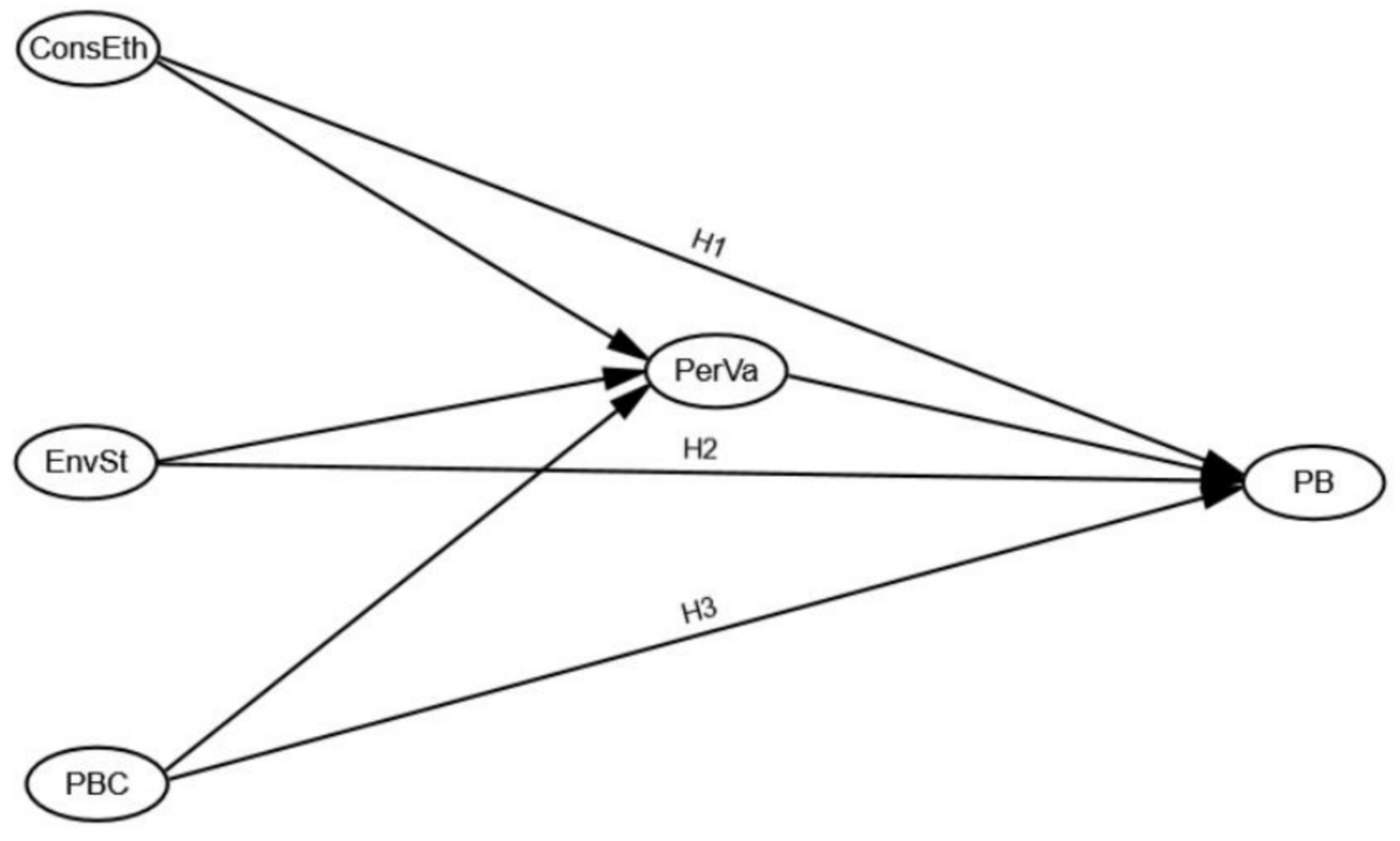}
\caption{Research model.}
\label{Fig1}
\end{figure}

\begin{figure}[!htbp]
\centering
\includegraphics[width=0.45\textwidth]{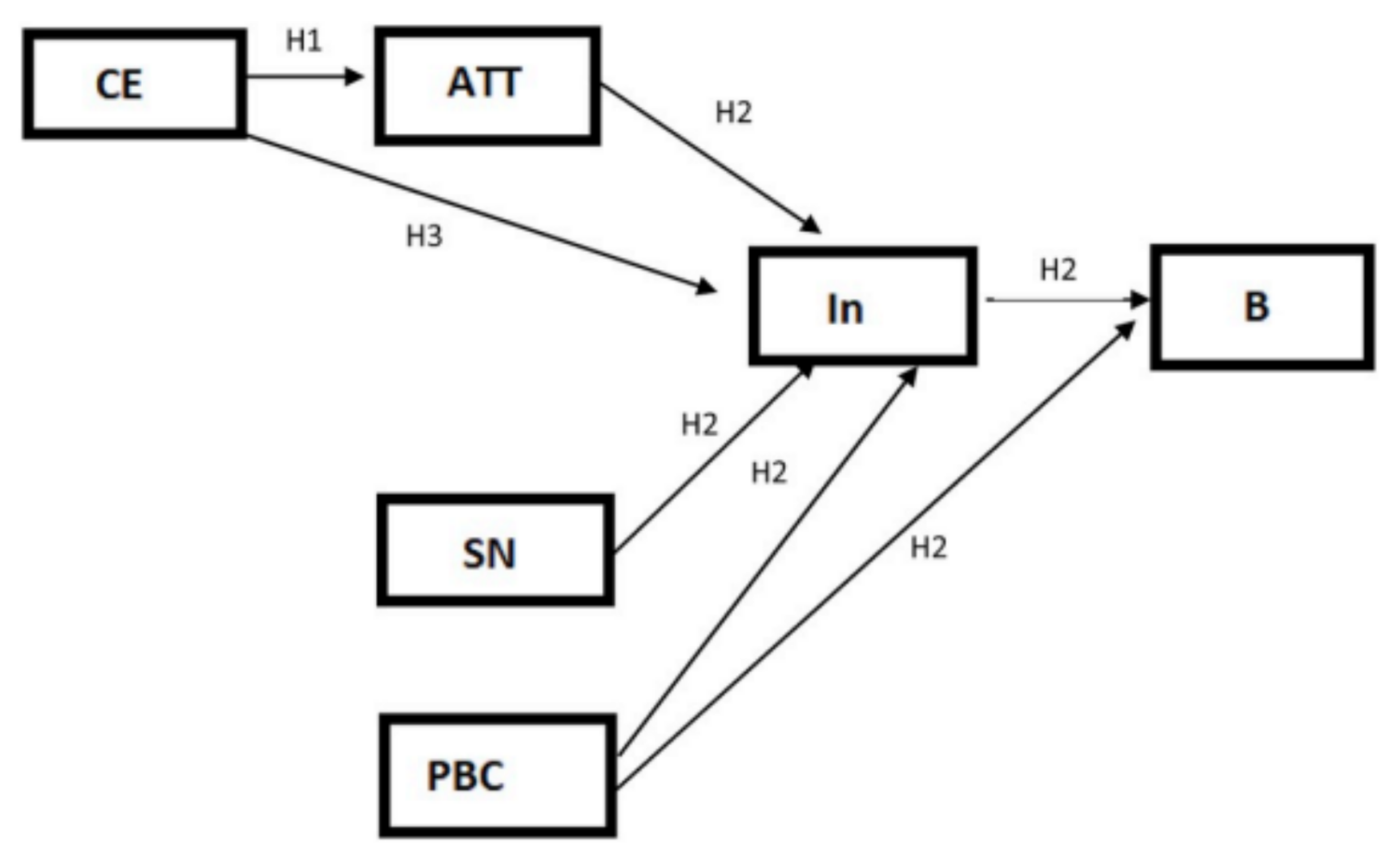}
\caption{Research model in \cite{authour7}.}
\label{Fig2}
\end{figure}

\begin{figure}[!htbp]
\centering
\includegraphics[width=0.45\textwidth]{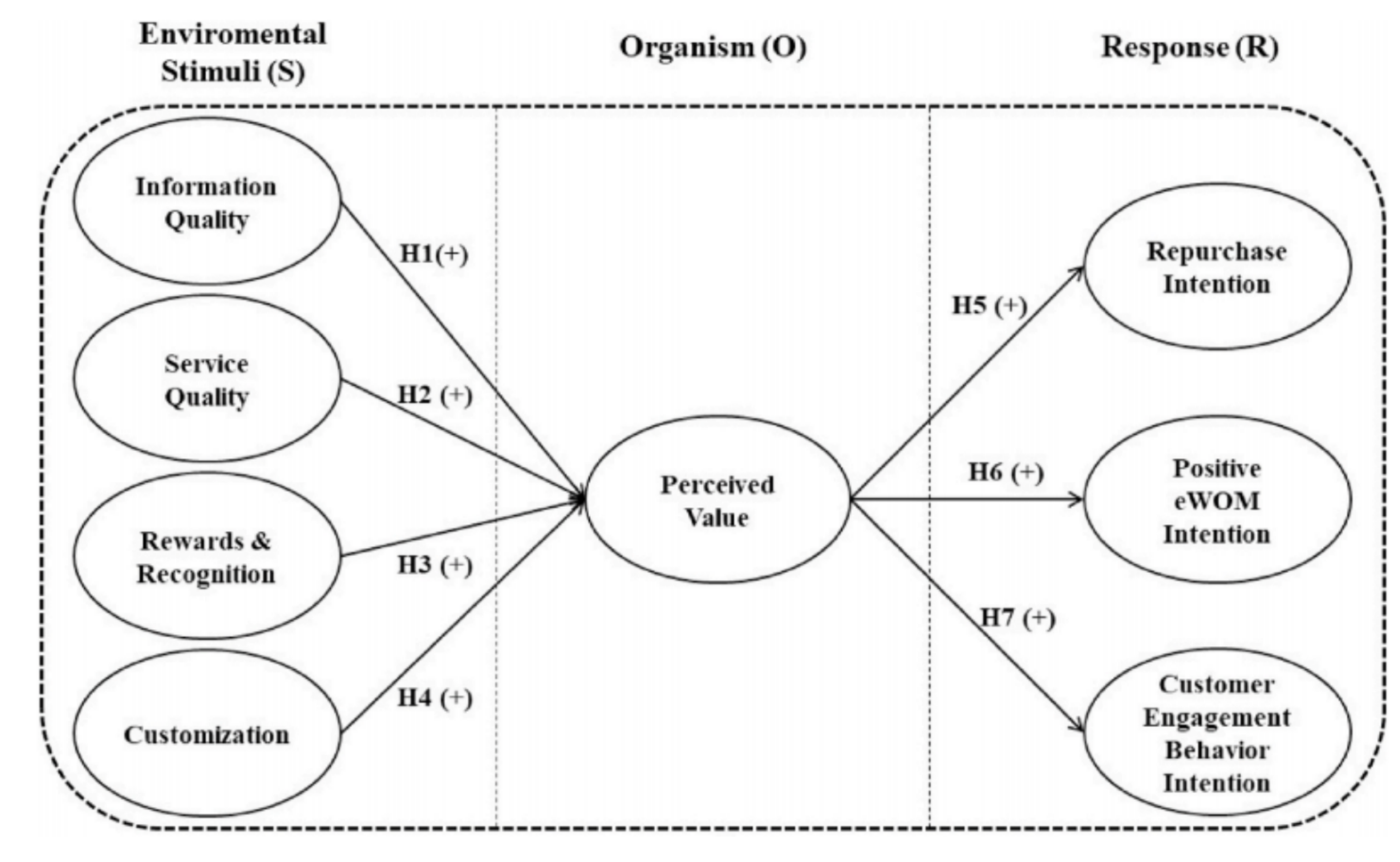}
\caption{Research model in \cite{authour8}.}
\label{Fig3}
\end{figure}

\begin{figure}[!htbp]
\centering
\includegraphics[width=0.45\textwidth]{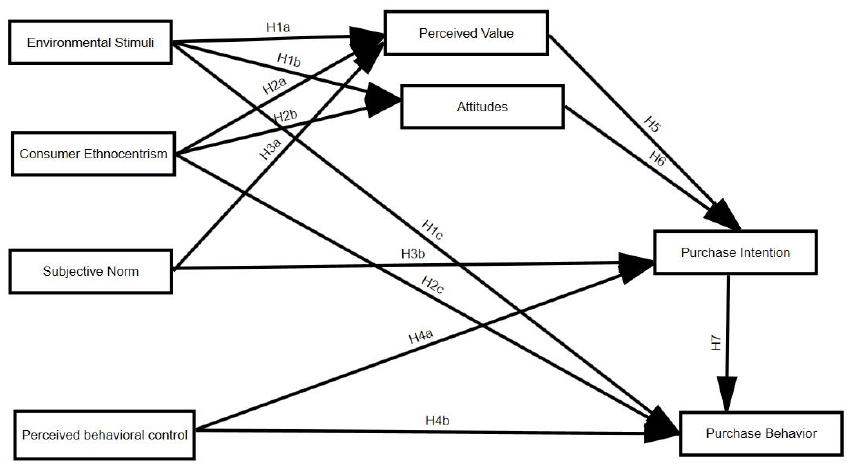}
\caption{Research model in \cite{authour9}.}
\label{Fig4}
\end{figure}

\begin{remark}
In Figure 1, ConsEth=Consumer ethnocentrism, EnvSt=Environmental stimuli, PBC=Perceived behavioral control, PerVa=Perceived value, PB=Purchase behavior. and the conceptual definitions and factors included in this model can be found in References \cite{authour7,authour8,authour9,authour10,authou11,authour12,authour13,authour14,authour15,authour16,authour17,authour18,authour19,authour20,authour21,authour22,authour23,authour24,authour25,authour26,authour27,authour28,authour29,authour30,authour31}.
\end{remark}

\subsection{Research hypotheses}
Based on the structural equation model (Figure 1) and relevant hypotheses from References \cite{authour7,authour8,authour9,authour10,authou11,authour12,authour13,authour14,authour15,authour16,authour17,authour18,authour19,authour20,authour21,authour22,authour23,authour24,authour25,authour26,authour27,authour28,authour29,authour32,authour33,authour34,authour35,authour36,authour37,authour38,authour39}, this study proposes the following six hypotheses.
\\
H1: Consumer ethnocentrism positively influences purchase behavior.\\
H2: Environmental stimuli (EnvSt) positively influence purchase behavior.\\
H3: Perceived behavioral control (PBC) positively influences purchase behavior.\\
H4:  Perceived value (PerVa) plays a partial mediating role in the effect of consumer ethnocentrism (ConsEth) on purchase behavior (PB).\\
H5: Perceived value (PerVa) plays a partial mediating role in the effect of environmental stimuli (EnvSt) on purchase behavior (PB).\\
H6:Perceived value (PerVa) plays a partial mediating role in the effect of  perceived behavioral control (PBC) on purchase behavior (PB).

\section{Main results}
\subsection{Descriptive statistical results}

  In October 2024, an online survey (ID: 284362158) was conducted via the Wenjuanxing platform(www.wjx.com) , targeting consumers of Wuliangye liquor aged 18 and above. Participation was restricted to individuals whose IP addresses were verified to be located within Sichuan Province. Among 654 participants, 519 responses met validity criteria.The valid respondent pool exhibited a close-to-equal gender ratio, with $45.47\% $ male participants compared to $54.53\%$ female counterparts, as detailed in Table 1. By age cohort, Generation Z (18-30 years) accounted for $35.44\%$, while the largest segment fell within the 30-40 age range ($54.14\%$), with individuals over 40 representing $10.4\%$. A majority ($75.34\%$) held bachelor's degrees as their highest educational attainment, with only $8.86\%$ possessing master's degrees or higher. Professionally, $75.53\%$ of the 519 valid respondents were ordinary office workers. In terms of income, $60.5\%$ were classified as high earners (monthly income exceeding 7000 yuan), while $30.25\%$ fell into the moderate-income bracket (4001-7000 yuan).

  \begin{table}[!htbp]
  \begin{center}
  \caption{ Sample description. }
                         \begin{tabular}{|c|c|c|c|c|c|c|c|}
                          \hline n=519   &  & N&Percentage\\
                          \hline   \multirow{2}{*}{Gender } &  Male  &236 &$45.47\%$\\
                          \cline{2-4}
                              & female  &283 &$54.53\%$  \\
                         \hline      \multirow{3}{*}{Age }&18-29(Generation Z)&184&	$35.44\%$\\
                         \cline{2-4}
                                      &  30-40&	281&	$54.14\%$\\
                           \cline{2-4}
                          & Over 40 years old&	54&$10.4\%$\\
                          \hline       \multirow{3}{*}{Education }&Bachelor degree&	391&$	75.34\%$ \\
                          \cline {2-4}
                                 &Master/PhD&	46&	$8.86\%$\\
                                    \cline {2-4}
                                    &Others	&82&$15.8\%$\\
                             \hline    \multirow{3}{*}{Income}&Low&	48	&$9.25\%$\\
                             \cline {2-4}
                                    &Medium&	157&$	30.25\%$\\
                             \cline {2-4}
                                      &High	&314&$	60.5\%$\\
                             \hline     \multirow{3}{*}{Work role}   &ordinary office workers	&392&$	75.53\%$\\
                             \cline {2-4}
                                    &Other occupations	&89&$	17.14\%$\\
                             \cline {2-4}
                             &Others	&38&$	7.32\%$\\
                           \hline
                           \end{tabular}
                           \end{center}
\end{table}

  Within the Wuliangye liquor portfolio, Wuliangye is positioned as a premium offering, whereas Wuliangchun, Wuliangol, and Wuliangju are categorized as mid- to low-tier products (Table 2). Survey findings revealed that Wuliangye liquor dominated online purchases at $63.39\%$, followed by Wuliangchun at $59.15\%$. This emphasizes that premium liquors like Wuliangye liquor serve as  the primary purchasing choice for consumers in Sichuan Province, while Wuliangchun, a mid-to-low-tier liquor,   also demonstrates significant market appeal. Table 2 reveals that 319 out of 415 surveyed purchasers of Wuliangye liquor opted for online transactions, highlighting the critical  importance  of Wuliangye's  online shopping platform. Given Generation Z's
established preference for online shopping, the enhancement of Wuliangye's network platform infrastructure is not only essential but also strategically vital to securing the brand's future growth within the evolving liquor industry.

\begin{table}[bp]
 \begin{center}
 \caption{ Purchase preferences for Wuliangye series liquor. }
                         \begin{tabular}{p{2cm}p{0.5cm}p{1.5cm}p{0.5cm}p{1.5cm}}
                          \hline  &\multirow{1}{*}{ Purchase}&&	  \multirow{1}{*}{Online purchase} \\
                          \hline   n=519&	N&	Percentage&	N&	Percentage\\
                          \hline   Wuliangye&	415	&$79.96\%$&	329&$	63.39\%$\\
                          \hline   Wuliangol &	277	&$53.37\%$&	217&$	41.81\%$\\
                          \hline Wuliangchun&	378 &$	72.83\%$&	307&$	59.15\%$\\
                          \hline Wuliangju& 	210&$	40.46\%$&	129&$	24.86\%$\\
                           \hline
                           \end{tabular}
                           \end{center}
\end{table}

 \subsection{Statistical approach and analysis results}
 The assessment methods utilized a 7-point Likert scale ranging from strongly disagree (1) to strongly agree (7), with detailed procedural specifications comprehensively outlined in Table 3. Furthermore, according to Table 3 and Nunnally's established standards \cite{authour40}, a Cronbach's alpha value of at least 0.7 or 0.6 is considered acceptable. This confirms the data's suitability for subsequent exploratory factor analysis (EFA) conducted using SPSS software. Table 4 demonstrates near-perfect fulfillment of psychometric criteria (KMO $> 0.7$; Bartlett's test significance $< 0.01$), establishing both the structural robustness of the dataset and its eligibility for EFA (Cudeck \cite{authour41}; Solis-Galvan \cite{authour42}). Finally, Table 5 reveals that nearly all measurement items achieved factor loadings exceeding 0.5, with primary dimensions approaching 1.0, while remaining values fell below 0.4 - confirming adequate discriminant validity.

  To further ensure the robustness of the measurement model, I computed the composite reliability (CR) and average variance extracted (AVE) for each construct using the following formulas \cite{authour43}:

 $$CR=\frac{(\sum\lambda_i)^2}{(\sum\lambda_i)^2+\sum \text{Var}(\epsilon_i)}, AVE=\frac{\sum\lambda_i^2}{\sum\lambda_i^2+\sum \text{Var}(\epsilon_i)}, $$
 where $\lambda_i$ denotes the standardized factor loading of item $i$,
  and $\text{Var}(\epsilon_i)$ is the error variance of item $i$. These indices are critical for assessing convergent validity and reliability of latent constructs \cite{authour43,authour44}.

The parameter estimation for the SEM was conducted using the Maximum Likelihood (ML) method, which maximizes the likelihood function under the assumption of multivariate normality:
$$\log L(\theta)=-\frac n2[\log|\sum(\theta)|+\text{tr}(S{\sum}^{-1}(\theta))+p\log(2\pi)],\eqno(3.1)$$
where $n$ is the sample size, $p$ is the number of observed variables,
$S$ is the sample covariance matrix,
$\sum(\theta)$ is the model-implied covariance matrix, and
$\theta$ is the vector of model parameters. This function is derived from the multivariate normal probability density function and measures the discrepancy between the sample covariance matrix and the model-implied covariance matrix \cite{authour31}.

For the readability of this article, the author is willing to demonstrate (3.1). Indeed, let $x_i$  be a
$p\times1$ vector of observed variables for the $i$-th observation, assumed to be independently and identically distributed from a multivariate normal distribution with mean vector $\mu(\theta)$  and covariance matrix $\sum(\theta)$. The probability density function for a single observation is:
$$f(x_i\mid \theta)=\frac1{(2\pi)^{p/2}|\sum(\theta)|^{1/2}}e^{-\frac12(x_i-\mu(\theta))^T\sum(\theta)^{-1}(x_i-\mu(\theta))}.$$
For a sample of $n$ independent observations, the joint likelihood function is:
$$L(\theta)=\prod_{i=1}^nf(x_i\mid \theta).$$
Taking the natural logarithm yields the log-likelihood:
$$\aligned
\log L(\theta)=&-\frac {np}2\log(2\pi)-\frac n2\log|\sum(\theta)|\\
&-\frac12\sum_{i=1}^n(x_i-\mu(\theta))^T{\sum}(\theta)^{-1}(x_i-\mu(\theta)).
\endaligned $$
In many SEM applications, the model is mean-structured such that $\mu(\theta)=0$. Under this condition, and using the trace identity
$\sum\limits_{i=1}^n x_i^TAx_i=\text{tr}(A\sum\limits_{i=1}^nx_i^Tx_i)$, the expression simplifies to:

$$\log L(\theta)=-\frac n2[p\log(2\pi)+\log|\sum(\theta)|+ \text{tr}({\sum}(\theta)^{-1}S)],$$
where $S=\frac1n \sum\limits_{i=1}^nx_i^Tx_i$ is the sample covariance matrix. This is equivalent to equation (3.1) up to an additive constant.

The fitting function to be minimized in ML estimation is:
$$F_{ML}=\log|\sum(\theta)|+\text{tr}(S{\sum}^{-1}(\theta))-\log|S|-p,\eqno(3.2)$$
where $p$ is the number of observed variables.  Minimizing $F_{ML}$
is equivalent to maximizing the likelihood function \cite{authour31}, as the two are related through a monotonic transformation. Specifically,
$$F_{ML}=-\frac2n\log L(\theta)-\log|S|-p\log(2\pi)+\text{constant}.$$

Since $\log|S|$ and $\log(2\pi)$ are constants with respect to $\theta$, minimizing $F_{ML}$
  is equivalent to maximizing $\log L(\theta)$. This establishes the theoretical foundation for using $F_{ML}$
  as the optimization criterion in maximum likelihood estimation.

To evaluate the overall model fit, I employed the following widely-used fit indices [36, 45]:

$$\chi^2/ df=\frac{\chi^2}{\text{degree of freedom}}, \text{RMSEA}=\sqrt{\frac{F_0}{df}},$$
$$\text{CFI}=1-\frac{\max(\chi^2_{\text{model}}-df_{\text{model}},0)}
{\max(\chi^2_{\text{null}}-df_{\text{null}},0)},$$
where $F_0$  is the population discrepancy function, and $\chi^2_{\text{null}}$ and $df_{\text{null}}$
  refer to the chi-square and degrees of freedom of the null model, respectively. These indices provide a comprehensive evaluation of model adequacy \cite{authour43}.

 Model fit was assessed using multiple indices, including Chi-square/DF, RMSEA, GFI, AGFI, TLI, NFI, CFI, PNFI, and PGFI, as summarized in Table 6 and Table 9. The criteria for acceptable fit follow established standards \cite{authour44,authour45}.

 Next, Figure 5 and Table 6 confirm satisfactory model fit. All constructs meet composite reliability (CR) benchmarks ($\geqslant0.7$) per Malhotra and Dash \cite{authour45}. Internal consistency validates per Fornell and Larcker \cite{authour44} with item loadings ¡Ý0.7. Average variance extracted (AVE) approaches/exceeds 0.5 (Fornell and Larcker \cite{authour44}), with consumer ethnocentrism showing strongest performance (Table 7). Discriminant validity is confirmed (Table 8), as AVE square roots generally exceed cross-construct correlations \cite{authour44}.

Finally,
Figure 6 and Table 9 confirm adequate model fit, validating the SEM's feasibility in revealing drivers of Wuliangye liquor purchases in Sichuan. Tables 10-11 show most path coefficients $>0.1$ (H3 exception). H1-H2 supported empirically; H3 rejected (p$>0.05$). In addition,
mediation analysis (Tables 12-14):

H4 (ConsEth$\rightarrow$PerVa$\rightarrow$PB): Significant total/direct effects but insignificant indirect effect $\rightarrow$ no mediation (rejected)

H5 (EnvSt$\rightarrow$PerVa$\rightarrow$PB): Significant total/indirect effects with insignificant direct effect$\rightarrow$full mediation (requires reformulation)

H6 (PBC$\rightarrow$PerVa$\rightarrow$PB): Insignificant total effect $\rightarrow$ unsupported

\begin{table}
\centering
\caption{ Reliability. }
                         \begin{tabular}{p{1.5cm} p{0.5cm}p{0.05cm}p{2cm}}
                         \hline  Variables&	NO.	& &	Cronbach's $\alpha$\\
                          \hline   \multirow{6}{*}{ConsEth} &  CE1 &     &\multirow{6}{*}{0.864}\\
                          \cline{2-3}
                          &CE3	 &\\
                            \cline{2-3}
                            &CE4	& \\
                            \cline{2-3}
                            &CE7	 &\\
                            \cline{2-3}
                            &CE9	 &\\
                            \cline{2-3}
                           & CE10	& \\
                           \hline   \multirow{3}{*}{EnvSt} &  ES1	&  &\multirow{3}{*}{0.678}\\
 \cline{2-3}
 &ES3	&  \\
 \cline{2-3}
 &ES4	&  \\
                          \hline   \multirow{3}{*}{PBC} &  PBC1  &  &\multirow{3}{*}{0.631}\\
                          \cline{2-3}
                              & PBC2  &     \\
                         \cline{2-3}
                              & PBC3 &    \\
                              \hline   \multirow{4}{*}{PerVa} &  PV1  &   &\multirow{4}{*}{0.717}\\
                          \cline{2-3}
                                         &PV2   & \\
                          \cline{2-3}
                                     &PV3 & \\
                            \cline{2-3}
                            &PV4 &  \\
                            \hline   \multirow{5}{*}{PB} &  PB1  &   &\multirow{5}{*}{0.836}\\
                          \cline{2-3}
                                         &PB2   & \\
                          \cline{2-3}
                                     &PB3 & \\
                            \cline{2-3}
                            &PB4 &  \\
                             \cline{2-3}
                            &PB5 &  \\
\hline
                           \end{tabular}
\end{table}
\begin{remark}All tables (Table 1-14) were adapted from SPSS Software or Amos software results. In Table 3, 	after conducting reliability and validity tests in the pre-test, certain items such as CE2, CE5, CE6, CE8, ES2, ES5-ES10, and PBC4 were eliminated and removed. The remaining selected valid variables and items are presented in Table 3. Here,
      CE1:Only those products that are unavailable in the China should be imported;
                                                   CE3:	 It is not right to purchase foreign products, because it puts Chinese people out of jobs;
                            CE4:	A real Chinese person should always buy the products made in China;
                            CE7:	It may cost me in the long run but I prefer to support Chinese products;
                            CE9:	We should buy from foreign countries only those products that we cannot obtain within our own country;
                            CE10:	Chinese consumers who purchase products made in other countries are
responsible for putting their fellow Chinese out of work;
   ES1:	This social commerce site of Wuliangye provides me with the precise information I need;
 ES3:	I think the information content provided by the Wuliangye social commerce site is reliable;
 ES4:	The Wuliangye social commerce site provides me with up-to-date information;
     PBC1: I have enough opportunity in making purchase decision;
                               PBC2:  I have the capacity to make a purchase decision;
                               PBC3: I have enough control in my purchase decision;
                               PV1:  Considering the money I pay for buying products on this social commerce site or  physical store, internet shopping or offline purchasing here represents  a good deal;
                                         PV2: Considering the effort I make in shopping at this social commerce site or physical store, internet shopping or offline purchasing here is worthwhile;
                                     PV3: Considering the risk involved in shopping at this social commerce site or  physical store, r                               internet shopping or offline purchasing here is of value;
                            PV4: Overall, internet shopping at this social commerce site or offline purchasing at this physical store delivers me good value;
                            PB1:	I am buying Wuliangye-series liquor regularly;
PB2:	In my shopping basket is regularly Wuliangye-series liquor;
PB3:	When I am buying liquor, I regularly choose Wuliangye-series liquor;
PB4:	In the past ten months, I have bought Wuliangye-series liquor;
PB5:	I have had many experiences of buying Wuliangye-series liquor.
\end{remark}

\begin{table}
\centering
\caption{Coefficients of  KMO  and Sig. for all the variables.}
\scriptsize
\begin{tabular}{|c|c|c|c|c|c|c|c|}
\hline  \bf Variables	&ConsEth	&EnvSt	&PBC	&PerVa	&PB\\
\hline  \bf KMO coefficient	&0.847	&0.658	&0.642	&0.753	&0.843\\
\hline  \bf   Sig.	&***	&***	&***	&***	&***\\
\hline
\end{tabular}
\end{table}

\begin{table} \caption{ Rotated Component Matrix. }
\centering
                         \begin{tabular}{p{0.5cm} p{0.5cm}p{0.05cm}p{0.5cm}p{0.5cm}p{0.5cm}p{0.5cm}}
                          \hline  \footnotesize\bf CE1	&0.641	&&&&		\\
                          \hline   \footnotesize\bf CE3	&0.803&&&&\\
                          \hline     \footnotesize\bf   CE4	&0.848&&&&\\
                           \hline     \footnotesize\bf   CE7&	0.843&&&&\\
                            \hline    \footnotesize \bf          CE9&	0.633&&&&\\
                            \hline     \footnotesize\bf          CE10&	0.804&&&&\\
                            \hline    \footnotesize \bf        ES1&&&			0.418&&\\
                            \hline     \footnotesize\bf        ES3&&&			0.505&&\\
                            \hline     \footnotesize\bf       ES4&&&			0.448&&\\
                            \hline    \footnotesize \bf       PBC1&&&&			0.771&\\
                            \hline     \footnotesize\bf        PBC2&&&&			0.607&\\
                            \hline     \footnotesize\bf        PBC3&&&&			0.66&\\
                             \hline     \footnotesize\bf        PV1&&&&&				0.808\\
                             \hline     \footnotesize\bf        PV2&&&&&				0.686   \\
                             \hline    \footnotesize \bf        PV3&&&&&				0.47\\
                             \hline     \footnotesize\bf        PV4&&&&&				0.628\\
                              \hline     \footnotesize\bf        PB1&&0.777&&&& 			 \\
                              \hline     \footnotesize\bf        PB2&&   0.76&&&& 			 \\
                              \hline     \footnotesize\bf        PB3&&0.668&&&& 			 \\
                              \hline     \footnotesize\bf        PB4&&0.655&&&& 			 \\
                              \hline     \footnotesize\bf        PB5&&0.738&&&& 			 \\
                                                           		                \hline
                           \end{tabular}
                           \end{table}

\begin{table}
\centering
\caption{ Model fit coefficients from CFA. }
                         \begin{tabular}{|c|c|c|c|c|c|c|c|}
                          \hline  \bf Item&	\bf Value&	\bf Standard&	\bf Meeting   the standard ?\\
                          \hline   Chi/DF&	2.727&	$< 5$	&Yes\\
                           \hline  RMSEA&	0.058&	$< 0.08$&	Yes\\
                            \hline   GFI&	0.919&$	> 0.9$&	Yes\\
                             \hline  AGFI&	0.9&$	> 0.9$&	Yes\\
                              \hline   TLI&	0.911&$	> 0.9	$&Yes\\
                               \hline   NFI	&0.9&$	> 0.9$&	Yes\\
                                \hline   CFI&	0.924&$	> 0.9$&	Yes\\
                                 \hline   PNFI&	0.755&$	> 0.5$&	Yes\\
                                  \hline    PGFI&	0.712&$	> 0.5$	&Yes\\
                                                   \hline
                           \end{tabular}
\end{table}

\begin{table}
\centering
\caption{ Convergent validity. }
                         \begin{tabular}{p{2.5cm} p{1cm}p{0.5cm}p{0.75cm}p{1cm}p{0.1cm}}
                         \hline  Path&	Estimate&		P&	CR&	AVE\\
                          \hline      CE1$\leftarrow$ ConsEth&	0.515&     &\multirow{6}{*}{0.8662	}&\multirow{6}{*}{0.5277}\\
                          \cline{1-3}
                           CE3	$\leftarrow$ ConsEth&	0.742&	 	***\\
                          \cline{1-3}
                           CE4	$\leftarrow$ ConsEth&	0.838&	 	***\\
                            \cline{1-3}
                           CE7	$\leftarrow$ ConsEth&	0.836&	 	***\\
                           \cline{1-3}
                           CE9	$\leftarrow$ ConsEth&	0.546&	 	***\\
                            \cline{1-3}
                           CE10	$\leftarrow$ ConsEth&	0.807&	 	***\\
                           \hline      ES1$\leftarrow$ EnvSt&	0.703&     &\multirow{3}{*}{0.6821	}&\multirow{3}{*}{0.4183}\\
                          \cline{1-3}
                          	ES3$\leftarrow$ EnvSt& 0.647&	 	***	\\
                          	 \cline{1-3}
                          	ES4$\leftarrow$ EnvSt& 0.585&	 	***	\\
                           \hline      PBC1$\leftarrow$ PBC &0.661&     &\multirow{3}{*}{0.6356}&\multirow{3}{*}{0.3699}\\
                          \cline{1-3}
                           PBC2$\leftarrow$ PBC& 0.631&		***\\
                           \cline{1-3}
                           PBC3$\leftarrow$ PBC& 0.524&	 ***\\
                            \hline      PV1$\leftarrow$ PerVa&	0.647&     &\multirow{3}{*}{0.7198}&\multirow{3}{*}{0.3913}\\
                          \cline{1-3}
                          	 PV2$\leftarrow$ PerVa& 0.626&	 	***	\\
                           \cline{1-3}
                          	 PV3$\leftarrow$ PerVa& 0.594&	 	***	\\
                          \cline{1-3}
                          	 PV4$\leftarrow$ PerVa& 0.634&	 	***	\\
                          \hline      PB1$\leftarrow$ PB &0.754&     &\multirow{5}{*}{0.8397}&\multirow{5}{*}{0.514}\\
                          \cline{1-3}
                           PB2$\leftarrow$ PB & 0.774&		***\\
                           \cline{1-3}
                    PB3$\leftarrow$ PB & 	0.669&		***\\
                    \cline{1-3}
                    PB4$\leftarrow$ PB & 0.594&		***\\
                    \cline{1-3}
                    PB5$\leftarrow$ PB & 0.776&		***\\
                    \hline
                           \end{tabular}
                           \end{table}

\begin{remark}In Table 7,   * represents $<0.1$, ** represents $<0.05$, and *** represents $<0.001$.\end{remark}

\begin{table}
\centering
\caption{ Discriminant validity. }
                         \begin{tabular}{|c|c|c|c|c|c|c|c|}
                          \hline  \it Variables	&PB&PerVa&PBC&EnvSt&ConsEth				\\
                          \hline  PB& \bf 0.717&&&&\\
                          \hline PerVa&	0.729**&\bf	0.626&&&\\
                          \hline     PBC&	0.507**&	0.648**&\bf	0.608&&\\		
\hline  EnvSt&	0.696**&	0.786**&	0.671**&\bf	0.647&\\	
\hline ConsEth&	0.295**&	0.205**&	-0.085*	&0.175**&	\bf0.726\\
\hline \it AVE&	0.514&	0.3913&	0.3699&	0.4183&	0.5277\\
                                                   \hline
                           \end{tabular}
                           \end{table}

\begin{remark} In Table 8, ** was significantly correlated at 0.001 level, and * was significantly correlated at  0.1 level. The value in bold on the upper right corner is the square root of AVE. For example, $0.717^2=0.514$.\end{remark}

\begin{table}
\centering
\caption{ Model Fit Summary of the SEM. }
                 \begin{tabular}{|c|c|c|c|c|c|c|c|}
                          \hline  \bf Item&	\bf Value&	\bf Standard&	\bf Meeting the standard ?\\
                          \hline   Chi/DF&	2.727&	$< 5$	&Yes\\
                            \hline   P&	0.000&	$< 0.05$	&Yes\\
                                 \hline   GFI&	0.919&$	> 0.9$&	Yes\\
                             \hline  AGFI&	0.9&$	> 0.9$&	Yes\\
                              \hline   TLI&	0.911&$	> 0.9	$&Yes\\
                               \hline   NFI	&0.9&$	> 0.9$&	Yes\\
                                \hline   CFI&	0.924&$	> 0.9$&	Yes\\
                                 \hline   PNFI&	0.755&$	> 0.5$&	Yes\\
                                  \hline    PCFI&	0.788&$	> 0.5$	&Yes\\
                                   \hline   RMSEA&	0.058&$	< 0.08$	&Yes\\
                                                   \hline
                           \end{tabular}
                           \end{table}

\begin{table}
\centering
\caption{ Regression weights. }
                         \begin{tabular}{p{2.6cm} p{0.9cm}p{0.5cm}p{0.5cm}p{0.4cm}p{0.2cm}p{0.9cm}}
                         \hline  Path&	Estimate&	S.E.&	C.R.&P&	Label\\
                          \hline      PerVa	$\leftarrow$	ConsEth&	0.124&	0.046&	2.357&	0.018&\\
                          \hline      PerVa	$\leftarrow$	EnvSt&	0.587&	0.126&	5.364&	***&\\
                           \hline      PerVa	$\leftarrow$	PBC	&	0.264&	0.136&	2.637&	0.008&\\
                           \hline      PB		$\leftarrow$	ConsEth&0.156&	0.068	&3.253&	0.001&	H1\\
                           \hline      PB		$\leftarrow$	EnvSt&0.301&	0.212&	2.658&	0.008&	H2\\
                           \hline      PB		$\leftarrow$	PBC	&0.034&	0.192&	0.392&	0.695&	H3\\
                       \hline      PB		$\leftarrow$	PerVa	&0.438&	0.179&	3.986&	***	 &	 \\
                  \hline
                           \end{tabular}
                           \end{table}

\begin{table}
\centering
\caption{  Results of Hypotheses (H1-H3). }
                         \begin{tabular}{p{0.2cm}p{2.3cm} p{0.7cm}p{1.95cm}p{1cm}p{0.05cm}}
                         \hline  No.& Path&	Path & Sig. &Supported\\
                                    &     &	coeff.& P-values  &	                     \\
                           \hline  H1&    PB		$\leftarrow$	ConsEth&0.156&	$0.001<0.05$&	yes \\
                           \hline   H2&   PB		$\leftarrow$	EnvSt&0.301	&$0.008<0.05$&	yes \\
                           \hline    H3&  PB		$\leftarrow$	PBC	&0.034&	$0.695>0.1$&	no \\
                  \hline
                           \end{tabular}
                           \end{table}

\begin{remark} In Table 11,  "Path coeff." means "Path coefficient", and "Sig. P-values" represents "Significance
 P-values".
 \end{remark}

\begin{table}
\centering
\caption{  Total effect, Direct effect  and Indirect effect of H4. }
                         \begin{tabular}{ |c||c|c|  }
                         \hline   \multicolumn{3}{|c|}{Estimate on H4 : ConsEth$\rightarrow$ PerVa$\rightarrow$ PB }  \\
                             Total effect	&Lower bound 	&Upper bound\\
   0.210	&0.110 	&0.308\\
                        Direct effect&	Lower bound& 	Upper bound\\
0.156&	0.051&	0.266  \\
Indirect effect&	Lower bound& 	Upper bound\\
  0.055	& -0.001&	0.155 \\
                                            \hline
                                                                             \end{tabular}
                           \end{table}

\begin{table}
\centering
\caption{  Total effect, Direct effect  and Indirect effect of H5. }
                         \begin{tabular}{ |c||c|c|  }
                                         \hline   \multicolumn{3}{|c|}{Estimate on  H5 : EnvSt$\rightarrow$PerVa$\rightarrow$PB}\\
                               Total effect	&Lower bound 	&Upper bound\\
  0.559	&0.291& 	0.819\\
                        Direct effect&	Lower bound& 	Upper bound\\
0.301&	-0.141 &	0.682   \\
Indirect effect&	Lower bound& 	Upper bound\\
    0.257	& 0.076&	0.662\\
                                                \hline
                           \end{tabular}
                           \end{table}

\begin{table}
\centering
\caption{  Total effect, Direct effect  and Indirect effect of H6. }
                         \begin{tabular}{|c||c|c|}
                                      \hline      \multicolumn{3}{|c|}{Estimate on  H6 : PBC$\rightarrow$PerVa$\rightarrow$PB}\\
                               Total effect	&Lower bound 	&Upper bound\\
  0.150&	-0.125& 	0.414\\
                        Direct effect&	Lower bound& 	Upper bound\\
0.034&	-0.226	&0.282  \\
Indirect effect&	Lower bound& Upper bound\\
      0.116	& -0.043&	0.344  \\
                                                \hline
                           \end{tabular}
                           \end{table}
  \begin{figure}[!htbp]
  	\centering
  	\includegraphics[width=0.45\textwidth]{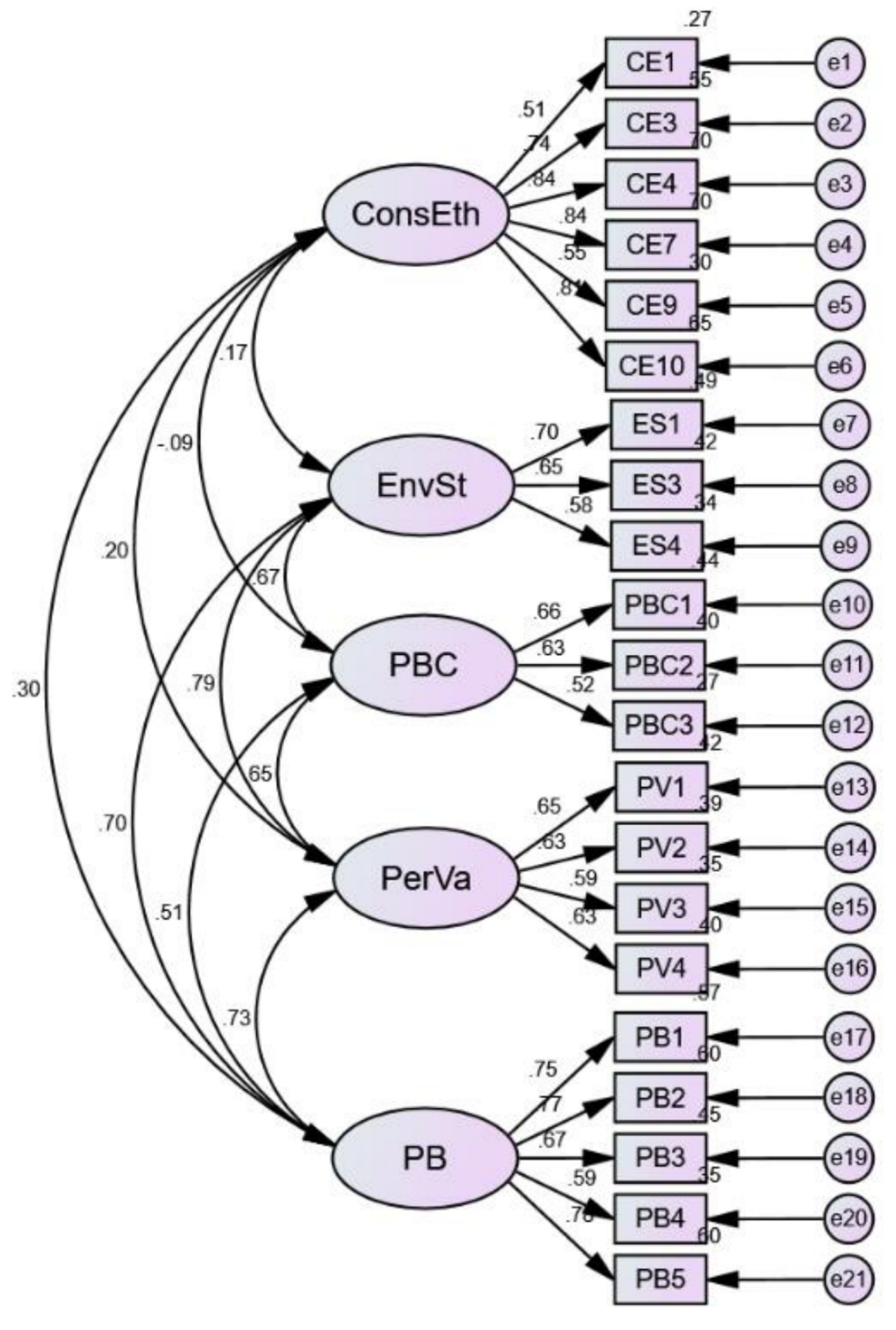}
  	\caption{ CFA.}
  	\label{Fig5}
  \end{figure}

  \begin{figure}[!htbp]
 	\centering
 	\includegraphics[width=0.45\textwidth]{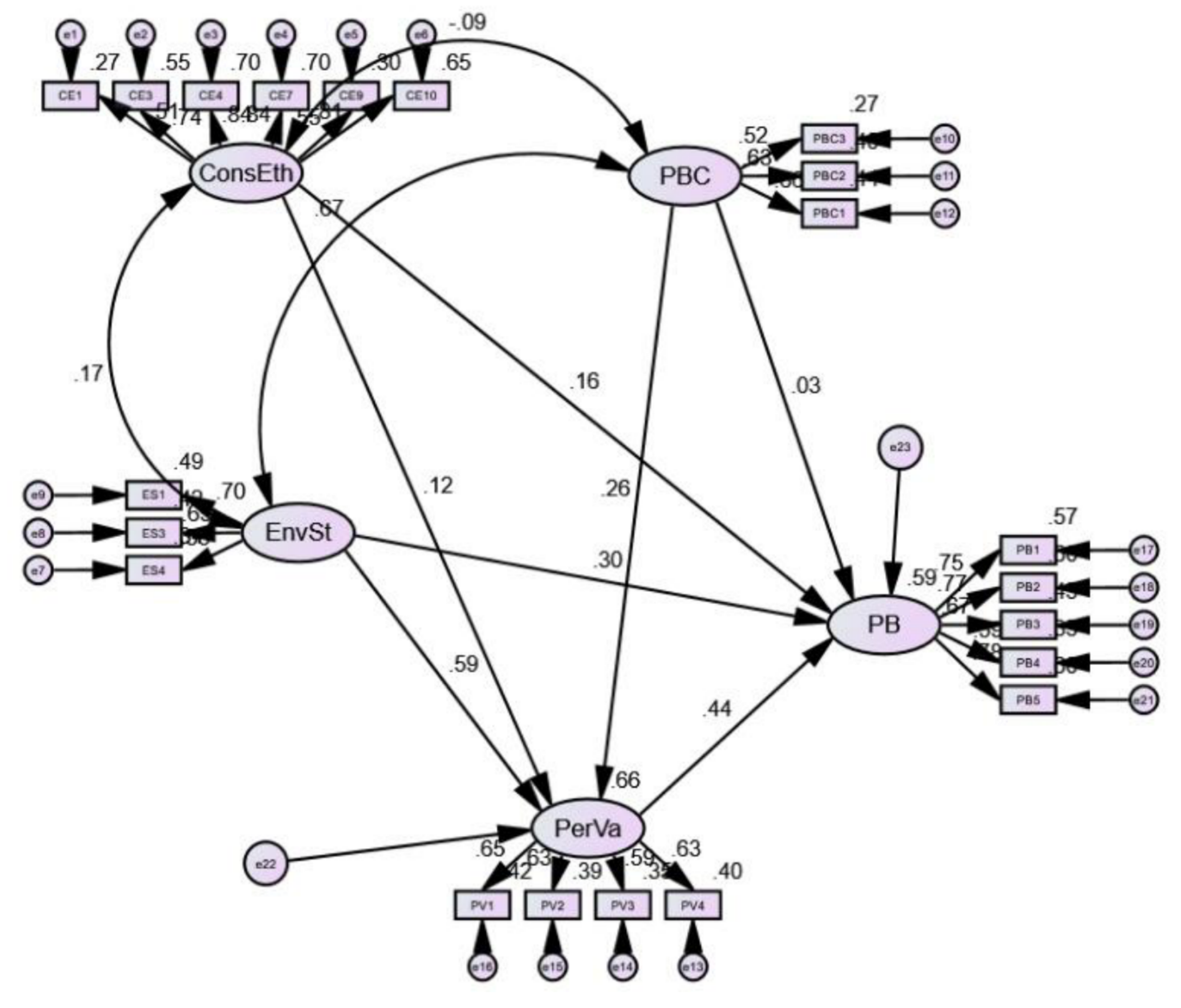}
 	\caption{SEM.}
 	\label{Fig5}
 \end{figure}

\begin{remark} This paper advances beyond Reference \cite{authour12}, which employed a TOPSIS-SEM-ANN hybrid model to analyze ISP consumer preferences in the Philippines, identifying SERVQUAL factors (assurance, responsiveness, empathy, data privacy) as drivers of service quality, satisfaction, and loyalty. While \cite{authour12} focused on ranking ISPs and predicting behavior through service quality attributes, this study introduces a superior framework by integrating ETPB and SOR theories into a unified model. It examines internal cognitive mechanisms (consumer ethnocentrism, perceived behavioral control) and external market stimuli (environmental cues) within Sichuan's socio-cultural context, specifically targeting premium liquor consumers. Methodologically, it explicitly models mediated effects (e.g., perceived value as a mediator), enabling deeper causal analysis of purchase behavior drivers, unlike \cite{authour12}'s direct relationships. Strategically, it provides actionable insights for Wuliangye's regional marketing through ethnocentrism leverage and e-commerce optimization, surpassing \cite{authour12}'s general ISP-focused approach. This paper thus offers a theoretically richer, context-specific, and strategically relevant framework for dissecting regional consumer behavior.
\end{remark}

\begin{remark}This study advances Reference \cite{authour13}, which employed a TOPSIS-SEM-ANN framework to evaluate Filipino ISP preferences using SERVQUAL dimensions, identifying assurance, responsiveness, empathy, and data privacy as key service quality drivers impacting satisfaction and loyalty. While \cite{authour13} provided valuable insights into utilitarian service evaluation, the present work introduces theoretical and contextual innovations. First, it integrates ETPB and SOR theories into a unified model, simultaneously examining internal cognitive factors (e.g., ethnocentrism, perceived control) and external stimuli (e.g., environmental cues), enabling a more holistic causal analysis of purchase behavior. Second, unlike \cite{authour13}'s general consumer focus, this research targets premium liquor consumption in Sichuan Province, offering culturally embedded, actionable insights. Third, it explicitly tests mediated pathways (e.g., perceived value as a mediator), revealing nuanced behavioral mechanisms absent in \cite{authour13}. Finally, the hybrid ETPB-SOR framework enhances explanatory power for complex, culturally specific decisions, positioning this work as a more theoretically integrated, context-sensitive, and mechanistically detailed contribution to consumer behavior research.
\end{remark}
\begin{remark}
This study surpasses Reference \cite{authour14}, which employed a Discrete Choice Experiment (DCE) and Latent Class Model (LCM) to analyze sustainable wine preferences among consumers in two second-tier Chinese cities (Chengdu and Xi'an). By identifying five consumer segments based on self-consumption and gift-giving contexts, \cite{authour14} highlighted the need for context- and region-specific marketing strategies. In contrast, this paper advances theoretical and methodological rigor by integrating the Extended Theory of Planned Behavior (ETPB) and Stimulus-Organism-Response (SOR) frameworks into a Structural Equation Modeling (SEM) framework. Unlike \cite{authour14}'s descriptive segmentation, SEM enables simultaneous analysis of latent cognitive processes (e.g., ethnocentrism, perceived control) and external stimuli (e.g., environmental cues), testing direct/mediated effects while accounting for measurement error. Tailored to Sichuan Province-a key market for Wuliangye-this study leverages localized data to uncover nuanced mediation pathways (e.g., perceived value) and delivers actionable, model-driven strategies (e.g., e-commerce optimization). Its context-specific, theoretically unified approach offers superior explanatory power for regional premium liquor purchase behavior compared to \cite{authour14}'s broad, preference-based segmentation.
\end{remark}
\section{Conclusions}
This study provides several key contributions to academia. Theoretically, it advances consumer behavior literature by successfully integrating two established theories, ETPB and SOR, into a cohesive framework that offers superior explanatory power for regional premium product purchases. Methodologically, it demonstrates the application of SEM to rigorously test complex mediated relationships within a specific socio-cultural context. Empirically, it yields novel insights into the triggering mechanisms of purchase behavior for a dominant Chinese liquor brand in its core market, highlighting the critical mediating role of perceived value and the dominant influence of environmental stimuli, which prior studies had not sufficiently uncovered.

 Structural equation modeling via Amos reveals environmental stimuli as the dominant factor, indirectly driving Wuliangye purchases through perceived value-contrary to literature claiming equal impacts from ethnocentrism, stimuli, and behavioral control. Consumer ethnocentrism directly boosted purchase intent, while perceived control showed negligible effects. Statistical analysis aligns Generation  Z's premium liquor preference with online purchasing frequency. Strategic imperatives include: (1) enhancing e-commerce infrastructure, (2) training specialized promotional staff, (3) leveraging cultural localization via ethnocentrism, and (4)Prioritizing research and development efforts on premium product lines within the Wuliangye Group.

 {color{blue}It is important to note that while the findings are derived from a region-specific context (Sichuan Province) and a particular brand (Wuliangye), the theoretical model (ETPB-SOR integration) and the methodological approach (SEM with mediation analysis) are broadly applicable to other regional markets or product categories. The mechanisms underlying consumer decision-making-such as the role of perceived value, environmental stimuli, and ethnocentrism-are universal across contexts, though their relative importance may vary \cite{authour15}. Therefore, the insights generated here can inform decision-making in other regional markets for premium products, provided that local cultural and economic factors are appropriately considered.}

%%%%%%%%%%%%%%%%%%%%%%%%%%%%%%%%%%%%%%%%%%%%%%%%%%%%%%
%          AI TOOLS, USE AND LOCATION
%%%%%%%%%%%%%%%%%%%%%%%%%%%%%%%%%%%%%%%%%%%%%%%%%%%%%%
%We follow COPE's guidelines and policies regarding the use of Artificial Intelligence (AI) tools. COPE Policy on AI tools can be found at https://publicationethics.org/cope-position-statements/ai-author.

%Authors using AI tools in the writing of a manuscript, production of images or graphical elements of the paper, or in the collection and analysis of data, must be transparent in disclosing in this section how the AI tool was used and which tool was used. Authors are fully responsible for the content of their manuscript, even those parts produced by an AI tool, and are thus liable for any breach of publication ethics. - COPE

%Disclosure instructions

%If there is nothing to disclose, there is no need to add a declaration, otherwise please declare.

%\section*{Use of AI tools declaration}
%The author(s) declare(s) they have used Artificial Intelligence (AI) tools in the creation of this article.
%AI tools used:
%How were the AI tools used?
%Where in the article is the information located?

\section*{Use of AI tools declaration}
The author  declares he has not used Artificial Intelligence (AI) tools in the creation of this article.

\section*{Acknowledgment (All sources of funding of the study must be disclosed)}

This research is funded
by  Chengdu Normal University Campus Project (CS22XMPY0201).The author extends sincere thanks to the editor and reviewers for their valuable comments, time, and effort invested in the review process. Without these insightful contributions, this paper would not have reached its current form. The author also thanks Prof. Xiaodi Li for his helpful suggestions.

\section*{Conflict of interest}

The author has no conflicts of interest to declare that are
relevant to the content of this article.

\end{document}